\documentclass[aps,amsmath,twocolumn,prb,showpacs,superscriptaddress,floatfix]{revtex4}

\usepackage{graphicx}
\usepackage{color}

\begin{document}

\preprint{}

\title{Dynamics of coherent polaritons in double-well systems}

\author{D. Sarchi}
\email[]{davide.sarchi@epfl.ch}
\affiliation{Institute of Theoretical Physics, Ecole Polytechnique F\'ed\'erale de Lausanne EPFL, CH-1015 Lausanne, Switzerland}
\author{I. Carusotto}
\affiliation{BEC-CNR-INFM and Dipartimento di Fisica, Universit\`a
di Trento, I-38050 Povo, Italy}
\author{M. Wouters}
\affiliation{Institute of Theoretical Physics, Ecole Polytechnique
F\'ed\'erale de Lausanne EPFL, CH-1015 Lausanne, Switzerland},
\affiliation{TFVS, Universiteit Antwerpen, Groenenborgerlaan 171,
2020 Antwerpen, Belgium},
\author{V. Savona}
\affiliation{Institute of Theoretical Physics, Ecole
Polytechnique F\'ed\'erale de Lausanne EPFL, CH-1015 Lausanne,
Switzerland}

\date{\today}

\begin{abstract}
We investigate the physics of coherent polaritons in a double-well configuration under a resonant pumping. For a continuous wave pump, bistability and self-pulsing regimes are identified as a function of the pump energy and intensity. The response to an additional probe pulse is characterized in the different cases and related to the Bogoliubov modes around the stationary state.
Under a pulsed pump, a crossover from Josephson-like oscillations to self-trapping is predicted for increasing pump intensity.
The accurateness of the effective two-mode model is assessed by comparing its predictions to a full solution of the non-equilibrium Gross-Pitaevskii equation.
\end{abstract}


\maketitle                   

\section{Introduction}
Semiconductor microcavities in the strong coupling regime are
particularly well suited to study the physics of dilute Bose
gases in a solid state context \cite{pol_rev,deveaud:rev,Ciuti_review,Szy_review}.
The elementary excitations of the system consist of polaritons, i.e. a superposition of a cavity photon and an exciton which at low excitation levels satisfy Bose statistics.
Their photonic component guarantees that a large degree of spatial coherence is maintained in spite of disorder effects, while the excitonic one provides strong mutual interactions.
Differently from most other examples of Bose gases such as liquid $^4$He and ultracold atoms, a polariton gas is an intrinsecally non-equilibrium system, whose properties can dramatically differ
from the corresponding ones of systems at thermodynamical equilibrium~\cite{iac_superfl,coherence,szy_coh,wouters07}.

Recent advances in the semiconductor fabrication technology, have
made it now possible to design polariton traps with a high
flexibility in the shape and the depth of the trapping potential
\cite{bayer99,dasbach01,loffler05,eldaif06,kaitouni06,bajoni07,BEC_pol_traps}.
From this perspective, double-well potentials show a particular
interest as they provide a way of investigating the well-known
Josephson effect~\cite{pita03,josephson_th,josephson_exp} in
completely new non-equilibrium regimes. Some preliminary work for
the case of non-resonantly pumped polariton condensates has
recently appeared in~\cite{wouters07}, while many authors have
considered similar effects in a variety of different optical
systems~\cite{lugiato_review,iaccouplcav,iacSHG}.

In the present paper, we will concentrate on the case of resonantly and coherently pumped double-well polariton traps obtained by lateral patterning of a planar microcavity as experimentally done in~\cite{eldaif06,kaitouni06}: such a configuration allows not only for selective addressing and diagnostics of the two spatial modes, but also for a relatively easy time-resolution of the Josephson dynamics on a picosecond scale.
The mean-field calculations of the present paper will be a crucial preliminary step in view of truly quantum effects~\cite{arnaud} that are expected to take place in such miniaturized systems whenever the Josephson charging energy for a single polariton becomes comparable to both the linewidths and the hopping energy. This regime is expected to be entered in the next generation of samples.

In Sec.~\ref{par:theory}, we introduce the effective two mode
model and we write the motion equations describing the time
dynamics of the polariton field amplitudes in each of the two
wells. The phase diagram and the different instability regimes
are analyzed in Sec.~\ref{par:phdiagr} for the case of a
continuous-wave pumping. The result of a numerical integration of
the dynamical equations of the two-mode model is presented in
Sec.~\ref{par:quasi-cw} for the case of a quasi-continuous wave
pump and in Sec.~\ref{par:pulsed} for the case of a pulsed pump:
optical bistability and self-pulsing phenomena take place in the
former case, Josephson oscillations and self-trapping in the
latter one. In Sec.~\ref{par:polaritons}, the predictions of the
two-mode model are compared to full numerical simulations of the
generalized Gross-Pitaevskii equation. The observability of all
the predicted features is verified using realistic parameters for
coupled polariton boxes. Conclusions are finally drawn in
Sec.~\ref{par:conclu}.

\section{The two-mode model}
\label{par:theory}

A widespread description of the Josephson dynamics in two-well system is based on an effective two-mode model~\cite{josephson_th}. In addition to the linear coupling $J$ and the cubic nonlinearity $g$,  a coherent pumping $F_{1,2}(t)$ and loss rates  $\gamma_{1,2}$ are to be included in order to describe the driven-dissipative nature of the present system.
The equations of motion for the mode amplitudes $\psi_{1,2}(t)$ then read~\cite{arnaud}:
\begin{eqnarray}
i\hbar\dot{\psi}_1&=&\left(\hbar\omega_1-i\frac{\gamma_1}{2}\right)\psi_1+g|\psi_1|^2\psi_1-J\psi_{2}+F_1(t), \label{eq:dyn2mod_1} \\
i\hbar\dot{\psi}_2&=&\left(\hbar\omega_2-i\frac{\gamma_2}{2}\right)\psi_2+g|\psi_2|^2\psi_2 -J\psi_{1}+F_2(t). \label{eq:dyn2mod_2}
\end{eqnarray}
In the absence of nonlinearity $g=0$ and pumping $F_{1,2}=0$, the eigenvalues of the linear equations are
\begin{eqnarray}
E_{+,-}=&&\frac{1}{2}\left(\hbar\omega_1-i\gamma_1+\hbar\omega_2-i\gamma_2\right)\nonumber
\\
&&\pm\frac{1}{2}\sqrt{\left[\hbar(\omega_1-\omega_2)-i(\gamma_1-\gamma_2)\right]^2+4J^2}:
\label{eq:sym_asym}
\end{eqnarray}
the linear coupling $J$ splits the unperturbed levels
$\hbar\omega_{1,2}$ into a pair of mixed eigenmodes with energies $E_{+,-}$.
For zero detuning, $\omega_1=\omega_2$, and equal loss rates $\gamma_1=\gamma_2$, the
energy splitting is $E_{-}-E_{+}=2J$, and the two corresponding
eigenmodes are a symmetric mode
$\psi_{+}=\psi_{s}=(\psi_1+\psi_2)/\sqrt{2}$ and an antisymmetric
mode $\psi_{-}=\psi_{a}=(\psi_1-\psi_2)/\sqrt{2}$.

Under a symmetric pump, $F_1(t)=F_2(t)$, only the symmetric mode $\psi_+$
is excited, while under an antisymmetric pump, $F_1(t)=-F_2(t)$ only
the antisymmetric mode $\psi_-$ is excited. Under
a pump acting only on one unperturbed mode that is $F_1(t)=F(t)$, $F_2=0$ both the eigenmodes are excited. In what follows, we will concentrate our attention on this last case.

This simple linear analysis is made richer by the presence of nonlinear terms.
Actually, a cubic nonlinearity has two main, and strictly related effects: it introduces
intensity-dependent shifts of the effective energy levels and can be responsible for dynamical instabilities.
Because of the nonlinearity, the effective eigenmodes of the system are no longer the symmetric and the antisymmetric ones. Therefore, to refer to the actual eigenmodes of the system, we prefer to adopt the notation ``+'' and ``-'' mode.

\subsection{The stationary state}

In the whole paper, we shall restrict our attention to the case
of a pump acting on a single mode, i.e. $F_j(t)=\delta_{j1}F(t)$:
this scheme is in fact the most interesting for applications and
is amenable to an almost fully analytical treatment.

We start by considering the steady state of the system under a continuous
monochromatic pump, $F_1(t)=e^{-i\omega t}F^{s}$, where the amplitudes of the two modes oscillate at the pump frequency
\begin{equation}
\psi_j(t)=e^{-i\omega t}\,\psi_j^s. \label{eq:rotwav}
\end{equation}
Substituting this ansatz into Eqs.~(\ref{eq:dyn2mod_1}-\ref{eq:dyn2mod_2}), the following stationarity equations are immediately obtained:
\begin{equation}
\left(\hbar\omega_j-\hbar\omega-i\frac{\gamma_j}{2}\right)\psi_j^s+g
n_j\psi_j^s -J\psi_{3-j}^s+\delta_{j1}F^s=0\,, \label{eq:stat}
\end{equation}
where $n_j\equiv|\psi_j^s|^2$ defines the stationary intensity in
the two modes. From Eqs.~(\ref{eq:stat}), it is straightforward to
show that the mode amplitude $\psi_1^s$ and the pump amplitude $F^s$
are uniquely determined for each given pair of values of the pump
energy $\hbar\omega$ and of the stationary amplitude $\psi^s_2$ in the
non-pumped mode $2$.
By rearranging Eqs.~(\ref{eq:stat}), one then obtains the final equations
\begin{eqnarray}
\psi_1^s&=&J^{-1}
\left[\left(\hbar\omega_2-\hbar\omega-i\frac{\gamma_2}{2}\right)\psi_2^s+g n_2\psi_2^s\right] \label{eq:phdiagr_1}\\
F^s&=&-\left(\hbar\omega_1-\hbar\omega-i\frac{\gamma_1}{2}\right)\psi_1^s-g
n_1\psi_1^s+J\psi_{2}^s, \label{eq:phdiagr_2}
\end{eqnarray}
from which the $F^s(\hbar\omega,n_2)$ and
$\psi_1^s(\hbar\omega,n_2)$ diagrams in the frequency-intensity
$(\hbar\omega,n_2)$-plane shown in Sec.~\ref{par:phdiagr} will be
obtained.

\subsection{Stability of the stationary solution}
\label{sec:stab}

The stability of the stationary solutions found in the previous section can be assessed evaluating the spectrum of small fluctuations around the stationary solution:
\begin{equation}
{\psi}_j(t)=e^{-i\omega t}\left(\psi_j^{s}+\delta\psi_j(t)\right), \label{eq:steady}
\end{equation}
By linearizing the motion equation Eqs.~(\ref{eq:dyn2mod_1}-\ref{eq:dyn2mod_2}) around the stationary solution, one obtains the following linear equations:
\begin{multline}
\frac{d\,\delta\psi_j}{dt}=\left(\hbar\omega_j-\hbar\omega-i\frac{\gamma_j}{2}\right)\,\delta\psi_j \\
+2g\,|\psi_j^s|^2\,\delta\psi_j
+g\left(\psi_j^s\right)^2\,\delta\psi_j^*-J\,\delta\psi_{3-j}.\label{eq:fluctdyn}
\end{multline}
Substituting in Eqs.~(\ref{eq:fluctdyn}) the time evolution
\begin{equation}
\delta\psi_j(t)=e^{-iE t/\hbar}U_j+e^{iE^* t/\hbar}V^*_j,
\label{eq:flfield}
\end{equation}
expressed in terms of the excitation energies $E$ and of the
fluctuation amplitudes $U_j$ and $V_j$, the problem reduces to
the secular equation
\begin{equation}
M\cdot \delta\Psi =E\, \delta\Psi \label{eq:eig},
\end{equation}
where we have introduced the vector $\delta\Psi=(U_1 V_1 U_2
V_2)^T$ and the matrix $M$ has the Bogoliubov form
\begin{widetext}
\begin{equation}
M=\left(\begin{array}{cccc} \hbar\tilde{\omega}_1-i\frac{\gamma_1}{2}+2 g n_1 & g\left.\psi_1^s\right.^2 & -J& 0\\
-g\left(\left.\psi_1^{s}\right.^*\right)^2 &-\hbar\tilde{\omega}_1-i\frac{\gamma_1}{2}-2 g n_1 & 0& J\\
-J& 0&  \hbar\tilde{\omega}_2-i\frac{\gamma_2}{2}+2 g n_2 & g\left.\psi_2^s\right.^2\\
0& J &-g\left(\left.\psi_2^{s}\right.^*\right)^2 & -
\hbar\tilde{\omega}_2-i\frac{\gamma_2}{2}-2 g n_2
\end{array}\right).
\label{eq:mtxbog}
\end{equation}
\end{widetext}
in terms of the frequencies $\tilde{\omega}_j=\omega_j-\omega$.

The resulting spectrum consists of four eigenvalues $E_{\alpha}$,
$\alpha=1,...,4$, corresponding to the normal modes
$\delta\Psi^{\alpha}$. As shown by Eq. (\ref{eq:flfield}), if the
imaginary parts of all the four energies are negative
$\mbox{Im}\{E_{\alpha}\}<0$, the fluctuation is damped and the
stationary solution is stable. On the other hand, if the
imaginary part of at least one eigenvalue is non-negative
$\mbox{Im}\{E_{\alpha}\}\geq 0$, the solution is unstable.

In this latter case, two situations are
possible: if $\mbox{Re}\{E_{\alpha}\}=0$ the solution is one-mode
(saddle-node) unstable (1M), while, if $\mbox{Re}\{E_{\alpha}\}>0$,
the solution is parametrically unstable (P). These two situations
will be discussed in detail in what follows.

\section{Continuous excitation: phase diagram and fluctuation spectrum}
\label{par:phdiagr}

\begin{figure}[ht]
\includegraphics[width=.48 \textwidth]{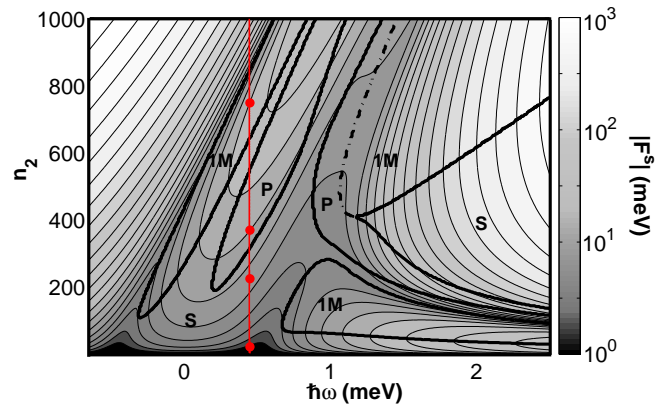}
\caption{Energy-intensity ($\hbar\omega$-$n_2$) diagram. The
colour scale corresponds to increasing values of $|F^s|$ in
logarithmic scale and the thin isolines are geometrically spaced
contour lines. The thick black line separates the stability ({\bf
S}) regions from instability ones: depending on their character,
these are marked by either {\bf 1M} (one-mode instability) or
{\bf P} (parametric instability). The dash-dot line separates a
one-mode instability region from a parametric instability one.
The dots on the vertical $\hbar\omega=0.45\,\textrm{meV}$ line
correspond to the pump parameters used later on in
Figs.~\ref{fig:TIMEFcont} and \ref{fig:FTFcont}. The system
parameters are inspired from the symmetric double-box polariton
traps discussed in Sec.~\ref{par:polaritons}, namely
$J=0.5~\mbox{meV}$, $\gamma_1=\gamma_2=\gamma=0.2~\mbox{meV}$,
$g=1.1~\times~10^{-3}~\mbox{meV}$, $\omega_1=\omega_2$.}
\label{fig:phdiagr}
\end{figure}

Under continuous monochromatic pumping, a contour plot of the
pump amplitude $F^s$ as a function of the pump energy
$\hbar\omega$ and of the intensity $n_2$ in the non-pumped mode
is readily obtained from
Eqs.~(\ref{eq:phdiagr_1}-\ref{eq:phdiagr_2}). An example of such
plot is shown in Fig.~\ref{fig:phdiagr} for the symmetric
$\omega_1=\omega_2$ case.

The phase diagram is determined from the stability of the
stationary state (\ref{eq:stat}). The black thick lines mark the
contours between the regions of stability and the regions of
instability, as obtained by solving the linearized problem
Eq.~(\ref{eq:mtxbog}). Here, regions of one-mode instability or
parametric instability are indicated by respectively {\bf 1M} and
{\bf P} following the definitions introduced in
Sec.~\ref{sec:stab}.

\begin{figure}[ht]
\includegraphics[width=.43 \textwidth]{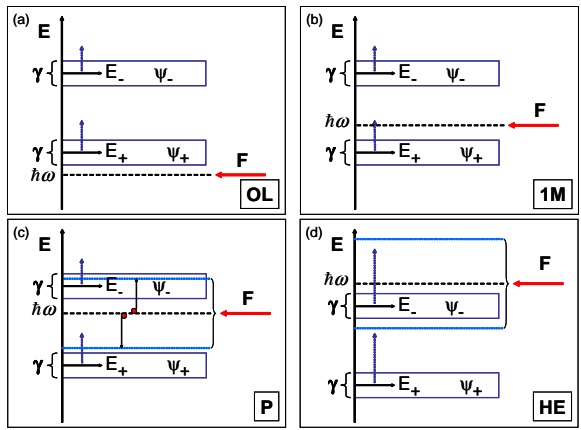}
\caption{Scheme of the effective energy levels $E_\pm$ of a two mode model under a continuous-wave pump. The vertical dashed arrows indicate the nonlinear blue-shift.
The horizontal solid arrows indicates the pump energy $\hbar\omega$ in the different regimes discussed in the text: (a) Optical limiter;
(b) One-mode instability; (c) Parametric instability; (d)
High-energy pumping.}\label{fig:scheme}
\end{figure}

For very small pump amplitudes, two resonances are clearly visible
in Fig.~\ref{fig:phdiagr}. As explained in Sec.~\ref{par:theory},
they corresponds to the $\psi_{\pm}$ eigenmodes of the linearly
coupled system and lye at exactly $\hbar\omega=\pm J$. The
linewidth has been taken smaller than the linear coupling
$\gamma/2<J$, so the corresponding lines are well distinct.

For increasing values of the pump amplitude, the resonances of the
system are modified by the nonlinearity (see the scheme in Fig.~\ref{fig:scheme}).
At moderate pump amplitudes for which $g\,n_{1,2}<J,\gamma$,
the main effect of the nonlinearity is a blue-shift of the two resonances.
On the other hand, several different instability mechanisms can take place
at larger pump amplitudes depending on the pump energy $\hbar\omega$.

\subsection{Optical Limiter}

For $\hbar\omega<-J$, the pump energy lies below the two energy
levels: this configuration, often called optical limiter (OL) in
the literature~\cite{NLO}, is stable for all pump intensities.
As shown in the sketch in Fig.~\ref{fig:scheme}(a), the effective energy levels are in fact pushed
even further off resonance from the pump by the nonlinear blue-shifts.

\subsection{Optical bistability}

For a pump energy just above the lower energy $+$ mode (i.e. $\hbar\omega\gtrsim -J$), the nonlinear shift is able to push the effective energy level into resonance with the pump  (see Fig.~\ref{fig:scheme}(b)) and give rise to a single mode (1M), saddle-node~\cite{cross} instability.
Analogous behaviour takes place for pump energies just above the higher energy $-$ mode (i.e. for $\hbar\omega\gtrsim J$).

The occurrence of regions of one-mode instability for energies higher than a mode resonance is a well studied subject in the general literature on instabilities~\cite{halebook}. Concerning nonlinear optical systems, it has been extensively studied both in the simplest case of single cavities~\cite{NLO}, as well as in more complex cases of coupled optical cavities \cite{lugiato_review,iaccouplcav} and OPOs \cite{baas_bistab,gippius04,wouters07a}.

\begin{figure}[ht]
\includegraphics[width=.43 \textwidth]{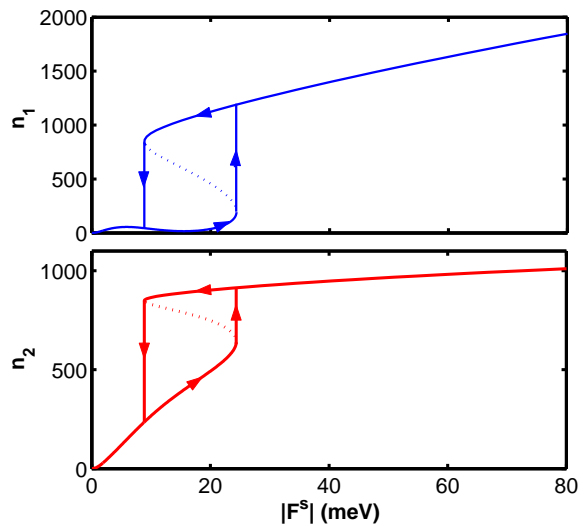}
\caption{Intensities $n_1$ (panel (a)) and $n_2$ (panel (b)) as a
function of the pump amplitude. Pump energy
$\hbar\omega=0.45~\mbox{meV}$. The arrows highlight the
hysteresis cycle due to optical bistability. Same system
parameters as in Fig.~\ref{fig:phdiagr}.} \label{fig:bistab}
\end{figure}

As a general feature~\cite{halebook}, one-mode instabilities of this kind often give rise to bistable behaviors, i.e. the coexistence of several  stable
solutions for the same values of the pump energy and amplitude.
An example of this behaviour is shown in Fig.~\ref{fig:bistab} where the dependence of the intensities $n_{1,2}$ on the pump amplitude is plotted for a pump energy  just above the lower resonance.

An hysteresis cycle is apparent: as the pump amplitude
$F$ increases from zero, the system moves along the lower branch of stable solutions until its end point is reached.
Only at this point the system jumps on the upper branch.
If the pump amplitude is then decreased, the system keeps moving along the upper branch of stable solution until its end point is reached, where it jumps back to the lower branch.

\subsection{Parametric instability}

For a pump energy between the two resonances, a parametric
instability appears. When the pump energy equals the average of
the effective energies $(E_+ + E_-)/2$, the parametric
process~\cite{NLO} sketched in Fig.~\ref{fig:scheme}(c) where the
pump field creates a signal $+$ and an idler $-$ fields becomes
resonant. This happens within the window
$\gamma_j<\hbar\omega<J+g n_2$, where the stationary solutions
are stable for small pump amplitudes, but become parametrically
unstable as soon as the parametric gain is able to overcome the
losses $gn_2 > \gamma_j$. Note that the effect of $n_1$ can be
neglected here, as in the considered energy window one has
$n_1\ll n_2$  (see Fig.~\ref{fig:n1vsn2}). In the dynamical
systems language, such an instability is called a Hopf
bifurcation~\cite{cross}.

The strong amplification of fluctuations around the stationary
solution eventually results in a self-pulsing dynamics where the
system keeps on oscillating for indefinite times. From a
different point of view, these oscillations can be seen as the
result of the interference between three fields at different
frequencies, i.e. the pump, signal, and idler fields of the
parametric oscillator~\cite{NLO,Ciuti_review}. This behavior will
be discussed in better detail in the next section.

\subsection{High-energy region}

For pump energy exceeding the energy of both resonances, several
instability regions are expected to appear as a consequence of
the complex interplay of single-mode 1M and parametric P
instabilities. Both effective energy levels eventually cross the
pump energy, as well as the parametric resonance
[Fig.~\ref{fig:scheme}(d)]. The diagram in Fig.~\ref{fig:phdiagr}
is therefore much richer in this window: for a given pump energy
$\hbar\omega$ and increasing values of $n_2$, three regions of
one-mode instability and three regions of parametric instability
can be identified, as well as a thin stability region in between
the first 1M and P regions.

\begin{figure}[ht]
\includegraphics[width=.43 \textwidth]{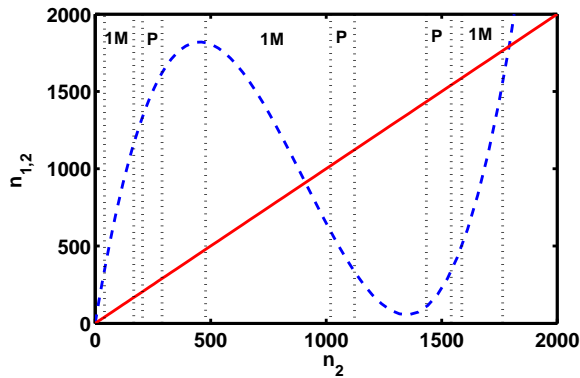}
\caption{Intensity $n_1$ as a function of $n_2$ for a high pump
energy $\hbar\omega=1.5~\mbox{meV}$ (dashed line). The solid line
corresponds to $n_2$ and is a guide for the eye in order to
identify the regions where $n_1\gg n_2$, $n_1\approx n_2$ or
$n_2\gg n_1$. Separation between instability regions of different
(1M or P) kinds  are indicated by the vertical dotted lines. Same
system parameters as in Fig.~\ref{fig:phdiagr}. }\label{fig:HEpop}
\end{figure}

\begin{figure}[ht]
\includegraphics[width=.48 \textwidth]{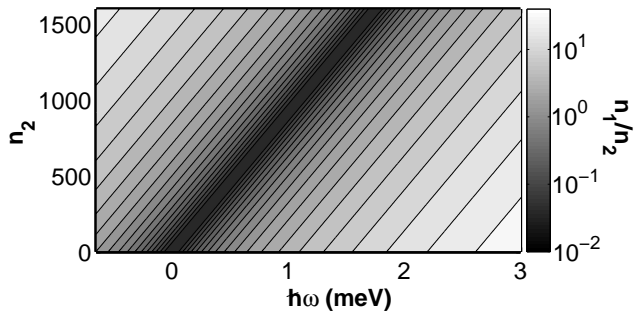}
\caption{Logarithmic gray-scale plot of the ratio $n_1/n_2$ as a
function of the pump energy and of the value $n_2$. Same
parameters as in Fig.~\ref{fig:phdiagr}. }\label{fig:n1vsn2}
\end{figure}

To understand the origin of the different regimes, the intensities $n_{1,2}$ are plotted in Fig.~\ref{fig:HEpop} as a function of $n_2$ for $\hbar\omega=1.5~\mbox{meV}$.
The behaviour is quite complex, yet can be analytically interpreted from Eq.~(\ref{eq:phdiagr_1}), which indeed gives
\begin{equation}
n_1=J^{-2}\left[(g n_2-\hbar\omega)^2+\frac{\gamma_2^2}{4}\right]n_2\,. \label{eq:n1vsn2}
\end{equation}
For a pump energy lying between the two blue-shifted resonances,
i.e. $-J+g n_2 <\hbar\omega<J+g n_2$, the intensity $n_2$ of the
non-pumped mode is larger than the intensity $n_1$ of the pumped
one, i.e. $n_2>n_1$. Conversely, for either lower
($\hbar\omega<-J+g n_2$) or larger ($\hbar\omega>J+g n_2$) pump
energies, the pumped mode has a larger intensity. This trend is
illustrated in the $n_1/n_2$ plot shown in Fig.~\ref{fig:n1vsn2}.

First we investigate the region around the stability tongue
extending at $\hbar\omega\gtrsim 1.5~\mbox{meV}$ for relatively
small values of $F$. While the energy of the pumped $1$ mode is
significantly blue-shifted by the nonlinear term, the non-pumped
$2$ mode remains almost empty (see Fig.~\ref{fig:HEpop}). The two
$1,2$ modes are then far in energy, so the effect of the linear
coupling is strongly suppressed. The nearby 1M instability region
then corresponds to the bistability loop for a pump close to
resonance with the higher energy mode, which in this region
basically coincides with the blue-shifted $1$ mode. The physics
is analogous for the second 1M instability region located just
above.

The third 1M instability at much larger pump amplitudes
corresponds to the opposite situation where the $2$ mode has been
shifted above $\hbar \omega$ and consequently $n_2\gg n_1$
according to (\ref{eq:n1vsn2}). The bistability loop then
involves the lower resonance, which in this regime basically
corresponds to the unperturbed $1$ mode.

We finally consider the intervals of parametric instability. The
first and the third intervals correspond to resonant scattering
processes taking place in a regime where the modes are effectively decoupled as
$n_{1} \gg n_{2}$ or $n_2 \gg n_1$, respectively.
In these regimes, the two $\pm$ eigenmodes essentially coincide with the $1,2$ modes.
The second interval corresponds instead to an intermediate regime where $n_1\approx n_2$, and the two $\pm$ eigenmodes are a superposition of both $1,2$ modes.

\subsection{The spectrum of fluctuations around the stationary solution}
\label{par:bogo}

\begin{figure}[ht]
\includegraphics[width=.43 \textwidth]{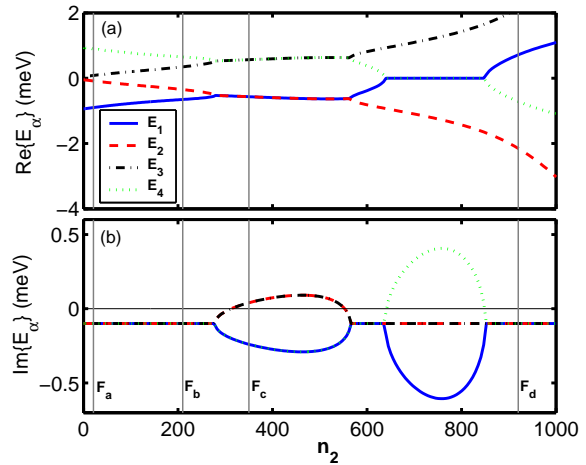}
\caption{Real (a) and imaginary (b) part of the linearized
spectrum around the steady state solution as a function of $n_2$.
Stable regions correspond to the imaginary part of the spectrum
being negative. Pump frequency $\hbar\omega=0.45~\mbox{meV}$. The
vertical lines indicate the $n_2$ values used in the panels (a-d)
of Figs.~\ref{fig:TIMEFcont} and \ref{fig:FTFcont}. Same system
parameters as in Fig.~\ref{fig:phdiagr}.} \label{fig:spectrum}
\end{figure}

The stability properties of the stationary solution discussed in the
previous section are further illustrated by looking at the eigenvalues
of the Bogoliubov linearized theory (\ref{eq:mtxbog}) of small fluctuations
around the stationary state~\cite{Ciuti_review,iac_superfl}.
These are plotted in Fig.~\ref{fig:spectrum} as a function of the intensity
$n_2$ of the non-pumped mode for the case of a pump energy $\hbar\omega$
chosen between the $\pm$ eigenmodes of the unloaded system.

In the linear $n_{1,2}\rightarrow 0$ regime, the frequencies and
damping rates tend to the ones
$\mbox{Re}\{E_{3,1}\}=-\mbox{Re}\{E_{4,2}\}=\pm J/2-\hbar \omega$,
$-\mbox{Im}\{E_{1,2,3,4}\}=\gamma/2$ of the unloaded system. As a
consequence of interactions, the frequency $\mbox{Re}\{E_{1,3}\}$
of the positive-weighted Bogoliubov modes are blue-shifted for
growing intensities, while the ones $E_{2,4}$ of the
negative-weighted ones are red-shifted: this makes them to
pairwise intersect at some value of the intensity. Here, each
pair collapses onto values
$\mbox{Re}\{E_{2(4)}\}=\mbox{Re}\{E_{1(3)}\}$ opposite in sign.
Correspondingly, the damping rate increases
$-\mbox{Im}\{E_1\}=-\mbox{Im}\{E_4\}>\gamma/2$ for two of them,
while it decreases $-\mbox{Im}\{E_2\}=-\mbox{Im}\{E_3\}<\gamma/2$
for the two others, possibly giving rise to a dynamical
instability, as in the case displayed in the figure. The fact
that frequencies of the modes involved in the instability are
non-zero is a signature of the parametric nature of the
instability.

For larger intensities, the frequencies split again within a
narrow intensity interval where the damping rate goes back to
$\gamma/2$, but another instability region occurs at even higher
intensities as a consequence of the intersection
$\mbox{Re}\{E_1\}=\mbox{Re}\{E_4\}$: while the the imaginary
parts of the $2,3$ modes stay unchanged, the ones of the $1,4$
modes are split and dynamical instability is signalled by one of
them becoming positive. Since the unstable mode has zero
frequency, the instability has the one-mode character typical of
optical bistability loops.

For even larger intensities, the stationary state becomes stable
again: because of the large blue- (red-) shift of the positive
(negative)-weighted Bogoliubov modes, no further intersections of
the mode frequencies can in fact occur.

\section{Continuous pump: bistability, self-pulsing, and response to a probe}
\label{par:quasi-cw}

The stationary states and the stability regions identified in the
previous section are a good starting point for the dynamical
study of the system that we carry out in the present section by
numerically solving Eqs. (\ref{eq:dyn2mod_1}-\ref{eq:dyn2mod_2}).
We first investigate the onset of the steady state when the pump
intensity is slowly increased in time to its asymptotic value.
Then we characterize the response of the system in its steady
state to an additional probe: this provides a simple and
effective way of measuring the frequencies and damping rates of
the Bogoliubov modes in the different regimes.

The quasi-continuous pump is assumed to have a smooth temporal profile of the form
\begin{equation}
F(t)=F_{max}\left(1-\frac{2}{1+e^{(t/\tau)^2}}\right)\,.
\label{eq:conpump}
\end{equation}
For very long switch-on time, the system evolves in a quasi-static way through a sequence of stable stationary states.

Once the system has got to its asymptotic stationary solution, a weak and short probe pulse is applied onto mode $1$. Its temporal shape is assumed to be a Gaussian
\begin{equation}
f_{g}(t) = f_{g}^0 e^{-(t-t_0)^2/\tau_g^2}\,, \label{eq:seed}
\end{equation}
its central frequency coincides with the one of the continuous
pump, and its duration $\tau_g$ is chosen to be short enough for
the pulse to encompass all the relevant spectral features, i.e.
all the 4 Bogoliubov modes shown in Fig.~\ref{fig:spectrum}.

\begin{figure}[htbp]
\includegraphics[width=.43 \textwidth]{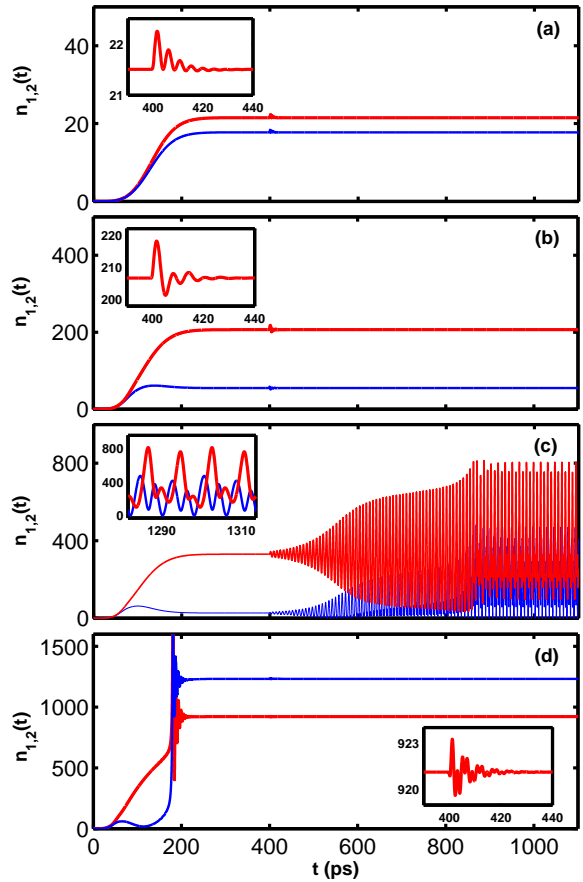}
\caption{Time-evolution of the $n_1(t)$ (dashed line) and
$n_2(t)$ (solid line) intensities for growing pump amplitudes
$F_{max}=1\,\mbox{meV}$ (a), $5\,\mbox{meV}$ (b) $8\,\mbox{meV}$
(c), $20\,\mbox{meV}$ (d). Pump energy
$\hbar\omega=0.45~\mbox{meV}$. Pump switch-on time
$\tau=100~\mbox{ps}$. Probe amplitude $f_g^0=10^{-2}\,F_{max}\ll
F_{max}$, probe duration $\tau_g=0.3\,\mbox{ps}\ll\pi\,\hbar/J$,
probe delay $t_0=400\,\mbox{ps}\gg \tau$. Same system parameters
as in Fig.~\ref{fig:phdiagr}. The pump parameters used for panels
(a-d) correspond to the red dots in Fig.~\ref{fig:phdiagr}.
}
\label{fig:TIMEFcont}
\end{figure}

\begin{figure}[htbp]
\includegraphics[width=.43 \textwidth]{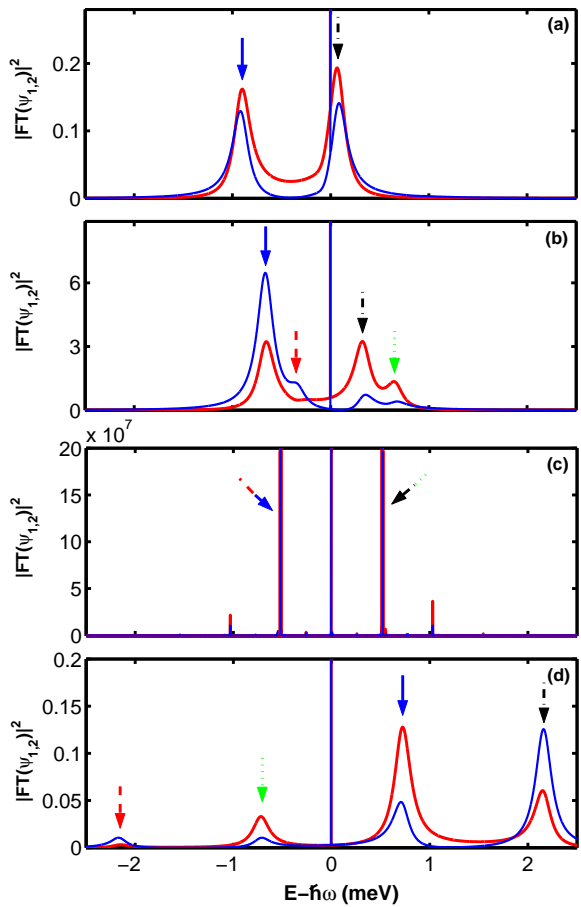}
\caption{Fourier transform spectra of $\psi_1$ (dashed line) and
$\psi_2$ (solid line) for growing pump amplitudes
$F_{max}=1,\,5,\,8,\,20~\mbox{meV}$. The central $\delta$-peak is
at the pump frequency. The arrows indicate the frequencies of the
Bogoliubov modes with the same colour code as in
Fig.~\ref{fig:spectrum}. Parameters of the (a-d) panels
correspond to the ones in Fig.~\ref{fig:TIMEFcont}. }
\label{fig:FTFcont}
\end{figure}

Numerical predictions for $n_{1,2}(t)$ are shown in
Fig.~\ref{fig:TIMEFcont} for the different regimes. The quasi-cw
pump energy $\hbar \omega$ is taken to be between the linear
resonance peaks and increasing values of the pump amplitude $F$
are chosen for the different panels (see dots in
Fig.~\ref{fig:phdiagr}). The corresponding spectra shown in
Fig.~\ref{fig:FTFcont} are obtained by Fourier transform of
$\psi_{1,2}(t)$. In order to eliminate complications due to the
switch-on dynamics, we have restricted the Fourier transform to
the temporal window following the arrival of the probe pulse. The
central $\delta$-peak corresponds to the pump frequency.

\subsection{Stable regime}

For small values of the asymptotic pump amplitude $F_{max}$
[Fig.~\ref{fig:TIMEFcont}(a) and (b)], the evolution during the
switch-on time smoothly leads the system to the asymptotic
stationary state. As this solution is stable, the response to the
probe pulse gets quickly damped within a time-scale of the order
of ten picosecond. While for very small intensities
(Fig.~\ref{fig:TIMEFcont}(a)) the response consists of damped
oscillations at a single frequency, for larger intensities
(Fig.~\ref{fig:TIMEFcont}(b)) relaxation is more complex and
involves interference of more frequencies.

This difference is apparent in the corresponding spectra shown in
Fig.~\ref{fig:FTFcont}(a,b) which are to be compared to the
Bogoliubov modes shown in Fig.~\ref{fig:spectrum}. Well inside
the stability region, only the positive-weighted Bogoliubov modes
with a significant $U$ component [see Eq.~(\ref{eq:flfield})] are
in fact visible. On the other hand, when the parametric
instability region is approached, the normal components $U$ are
significant for all modes and all the four frequencies become
then visible in the spectrum (see Fig.~\ref{fig:FTFcont}(b)). As
usual, the finite linewidth of the peaks is fixed by the finite
and negative imaginary part of the Bogoliubov modes, i.e. by
their damping rate. In the present case, this is the same for all
Bogoliubov modes.

\subsection{Parametric instability}

The physics is richer in Figs.~\ref{fig:TIMEFcont}(c) and
\ref{fig:FTFcont}(c) where a larger pump amplitude $F_{max}=8$
meV is considered: in this case, the asymptotic stationary state
is in fact parametrically unstable. As the pump intensity is
increased very smoothly, the system adiabatically follows the
stationary state at the instantaneous value of $F$ even in the
unstable region. However, the arrival of the probe pulse speeds
up the onset of the instability: the perturbation that it induces
is quickly amplified until the system gets to the self-pulsing
regime where undamped periodic oscillations take place for
indefinite time. Their frequency is close to the one of the
linear Bogoliubov mode getting unstable. Correspondingly, two
$\delta$-peaks appear in the spectrum shown
Fig.~\ref{fig:FTFcont}(c) at energies $E-\hbar\omega =\pm
0.52~\mbox{meV}$. The weaker $\delta$-peaks at harmonic
frequencies contribute to the quite complex waveform of the
self-pulsing oscillations in time shown in
Fig.~\ref{fig:TIMEFcont}(c).

\subsection{One-mode instability}

To highlight hysteresis phenomena, we now choose an asymptotic
value of the pump amplitude above the jump-up threshold of the bistability loop shown
in Fig.~\ref{fig:bistab}.
The switch-on then does not take place in a smooth way and the
system has to perform a sudden jump when the end-point of the lower
branch is reached: among the many complex behaviour that may take place~\cite{wouters07a},
the behaviour of the present system is the simplest: it jumps to the higher-intensity
branch of the bistable loop.
Once these upper branch is reached, the system is stable again and the response to
the probe pulse is qualitatively similar to the case of panels (a,b).
The only difference is the higher frequency of the oscillations, and the presence of
three different excitation frequencies which contribute to the complex relaxation dynamics:
one negative-weighted Bogoliubov mode has in fact a significant $U$ component.

\section{Pulsed excitation: Josephson oscillations and self-trapping}
\label{par:pulsed}

\begin{figure}[ht]
\includegraphics[width=.43 \textwidth]{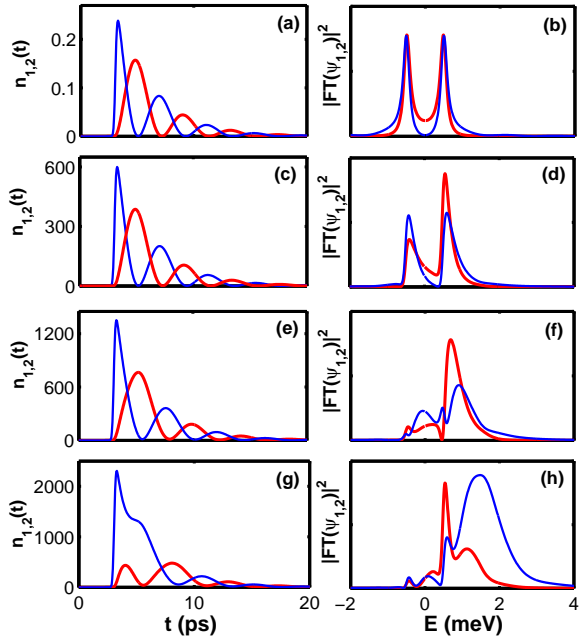}
\caption{Intensity dynamics (a,c,e,g) and corresponding Fourier
spectra (b,d,f,h) under a Gaussian pulsed pump (Eq.
(\ref{eq:pulsedF})) at time $t_p=3\,\mbox{ps}$ of duration
$\tau_p=0.2\,\mbox{ps}$. Peak pump amplitude
$F_{max}=1~\mbox{meV}$ (a,b), $50~\mbox{meV}$ (c,d),
$75~\mbox{meV}$ (e,f) and $100~\mbox{meV}$ (g,h). Blue (red)
lines corresponds to the $1$ ($2$) modes. Spectra are obtained by
Fourier transforming the signals in the whole time interval. Same
system parameters as in Fig.~\ref{fig:phdiagr}.}
\label{fig:pulsed}
\end{figure}

After having investigated the behaviour of the system under a
continuous pump, it is now interesting to look at the case where
only a pulsed pump is applied to the system. Specifically, we
numerically solve Eqs. (\ref{eq:dyn2mod_1}-\ref{eq:dyn2mod_2})
using a Gaussian temporal profile for the pump
\begin{equation}
F(t) = F_{max}\, e^{-(t-t_p)^2/\tau_{p}^2}\,. \label{eq:pulsedF}
\end{equation}
The duration $\tau_p$ of the pump pulse is taken to be very short as compared to all time scales of the system dynamics: the system is then almost istantaneously excited by a sudden kick, and then let evolve and relax without any further pumping.
The results are summarized in Fig.~\ref{fig:pulsed}(a,c,e,g) in the time domain, while the corresponding Fourier spectra are shown in Fig.~\ref{fig:pulsed}(b,d,f,h).

For low pump amplitudes in the linear regime, the intensities in the two wells manifest
Josephson-like oscillations [Fig.~\ref{fig:pulsed}(a)].
The pump pulse creates in fact a localized excitation in the $1$ mode, which is a superposition
of the symmetric and anti-symmetric eigenmodes of the system.
Because of their energy splitting, the system shows complete Josephson oscillations with a period $\pi\hbar/J$, which then damp out at a rate $\gamma$ under the effect of losses.
Correspondingly, the Fourier spectrum shown in Fig.~\ref{fig:pulsed}(b) is characterized by
a pair of resonance peaks split by $2J$.

For stronger pump amplitudes, the instantaneous intensity in the
system increases, and eventually results in significant nonlinear
effects. While the time evolution of the intensities
[Fig.~\ref{fig:pulsed}(c)-(e)] does not appear to be
significantly modified, nonlinear effects are visible in the
spectra of Fig.~\ref{fig:pulsed}(d)-(f) already at moderate
intensities as a global blue-shift of the spectrum and a
significant increase of the width of the peaks. In particular, in
Fig.~\ref{fig:pulsed}(f), note how the blue-shift of the spectrum
relative to the field $\psi_1$ is larger than the blue-shift of
the spectrum relative to $\psi_2$: this is due to the fact that
the intensity $n_1$ is at short times much larger than the
intensity $n_2$. In the Fourier spectrum of $\psi_1$, we also
clearly recognize two peaks at $J=\pm 0.5\,\textrm{meV}$
corresponding to the eigenfrequencies of the linear dynamics that
is recovered at long times once the intensities have dropped to
small values. Similar peaks also contribute to the Fourier
spectrum of $\psi_2$, but are hardly visible in the figure, their
weak intensity being hidden by the tails of the main peaks.

The appearance of peaks at frequencies characteristic of the nonlinear regime is a precursor of the
self-trapping regime that appears for stronger pump amplitudes: in this case, the nonlinear effects are in fact dominant in determining both the spectrum and the time evolution of the intensities [Fig.~\ref{fig:pulsed}(g)-(h)].

In the early stages of the evolution when the intensity is the largest, nonlinear effects dramatically suppress the amplitude of Josephson oscillation: most of the intensity is in fact {\em self-trapped} in the mode $1$ and the intensity of mode $2$ oscillates around a much smaller value.
As time goes on, the total intensity slowly drops under the effect of losses and eventually complete Josephson oscillations are recovered: the transition between the self-trapping regime and Josephson oscillations can be located in the vicinity of the time when the total intensity equals the critical density $n_1+n_2=n^c_{tot}=4 J/g$ of equilibrium Josephson systems~\cite{josephson_th}. The presence of losses is only responsible for a small shift of the critical point.

Consequences of this physics can be observed also in
Fig.~\ref{fig:pulsed}(h): both spectra show in fact two broad
peaks centered at high energies (i.e. at $E\simeq 0$ and $E\simeq
1.5~\mbox{meV}$), which represent the two resonances of the
system in the self-trapping regime, and two lower energy peaks at
$E=\pm J= \pm 0.5\,\textrm{meV}$, representing the frequencies of
the Josephson oscillations. These two latter peaks are asymmetric
[differently from the pure linear regime displayed in
Fig.~\ref{fig:pulsed}(b)] because the dynamics has been modified
by the occurrence of the self-trapping regime.

\section{Microcavity polariton boxes}
\label{par:polaritons}

In this last section we show how the parameters of the two-mode model can be evaluated
from the microscopic structure of a specific physical system.
On one hand we demonstrate that all the physics discussed in the previous sections can actually be observed in realistic systems, on the other hand we confirm the quantitative validity of the predictions of the two-mode model by comparing them to a full numerical integration of
the generalized Gross-Pitaevskii equation for polaritons~\cite{Ciuti_review,iac_superfl}.

Although many other configurations based on e.g. coupled DBR cavities are available to
study Josephson-like effects in an optical context \cite{lugiato_review,iaccouplcav,carole,trombettoni},
our attention will be concentrated on the specific case of double-well polariton traps.
Such a system was recently realized~\cite{eldaif06,kaitouni06} and combines the strong
nonlinearity due to the excitonic component of the polariton to the possibility of a micron-scale
spatial confinement by laterally patterning the thickness of the cavity layer.
From the point of view of Josephson physics, this geometry is very attractive as it allows
independent collection of light emitted from the two boxes and preserves the signal from being
covered by the incident laser field.

\subsection{From the non-equilibrium GPE to the two mode model}

The dynamics of the macroscopic polariton field $\Psi({\bf r},t)$ is described at mean-field level by a non-equilibrium generalization of the Gross-Pitaevskii equation (GPE) of the form~\cite{Ciuti_review,iac_superfl} :
\begin{multline}
i\hbar\,\frac{d}{dt}{\Psi}\left({\bf
r},t\right)=\left(-\frac{\hbar^2\nabla^2}{2m_p}+U_{ext}\left({\bf
r}\right)-i\frac{\gamma}{2}\right)\Psi\left({\bf r},t\right) \\
+ v\,\left| \Psi\left({\bf r},t\right)\right|^2\Psi \left({\bf
r},t\right)+f\left({\bf r},t\right)\,. \label{eq:gpe}
\end{multline}
$m_p$ is the effective mass of the lower polariton, $\gamma$
is the decay rate, $U_{ext}$ is the trapping potential, $v$ is the
effective polariton mutual
interaction~\cite{rochat00,ben01,okumura02} and $f({\bf r},t)$ is the amplitude of the coherent pump field.

\begin{figure}[ht]
\includegraphics[width=.43 \textwidth]{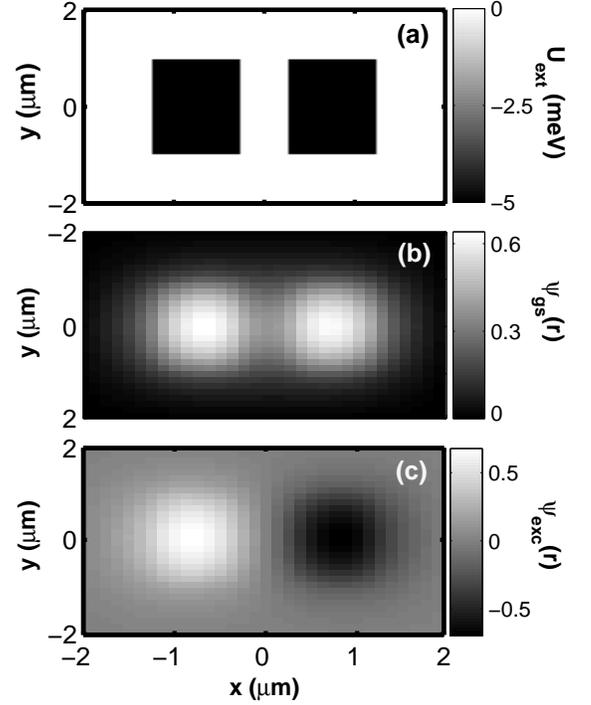}
\caption{Trapping potential $U_{ext}$ (Eq.~(\ref{eq:potwells}) due
to two adjacent rectangular wells (a), and the fundamental
$\phi_{gs}({\bf r})$ (b) and first excited $\phi_{exc}({\bf r})$ (c)
modes of the corresponding equilibrium GPE (\ref{eq:statgpe}). The
spatial profiles are displayed in grey tones.} \label{fig:potpsi}
\end{figure}

In this paper, we consider the case of a trapping potential
$U_{ext}({\bf r})$ formed by two adjacent wells, as displayed in
Fig.~\ref{fig:potpsi}(a):
\begin{multline}
U_{ext}=-U_0\,\theta\left(L_y-\left|y\right|\right)\left[\theta\left(L_x+{\delta}/{2}-x\right)\,
\theta\left(x-{\delta}/{2}\right)\right. \\ +\left.\theta\left(L_x+{\delta}/{2}+x\right)
\,\theta\left(-x-{\delta}/{2}\right)\right]\,.
\label{eq:potwells}
\end{multline}
For this potential, the fundamental mode of energy $E_{gs}$ is
described by an eigenfunction $\phi_{gs}({\bf r})$ which is
symmetric in the two wells (see Fig.~\ref{fig:potpsi}(b)), while
the first excited mode of energy $E_{exc}$ corresponds to an
antisymmetric eigenfunction $\phi_{exc}({\bf r})$ (see
Fig.~\ref{fig:potpsi}(c)). Without loss of generality, both these
functions can be taken real.

The pump field is considered to be monochromatic, with a gaussian spatial profile
\begin{equation}
f({\bf r},t)=2\pi\,\sigma^2\,f_0\,e^{-|{\bf r}-{\bf r_1}|^2/\sigma^2}\,e^{-i\omega t}
\end{equation}
centered on the first well at ${\bf r_1}=((L_x+\delta)/2,L_y/2)^t$.
Provided the pump energy is close to $E_{gs}$ and $E_{exc}$ and all other excited modes of the trapping potential are at much higher energies, the coherent field $\Psi({\bf r},t)$ can be safely written in the two-mode limit~\cite{josephson_th} as a time-dependent superposition of the two lower energy modes $\phi_{gs}({\bf r})$ and  $\phi_{exc}({\bf r})$ only.

For the present case, it is useful to write the superposition in the form
\begin{equation}
\Psi\left({\bf r},t\right)=\psi_1\left(t\right) \phi_1\left({\bf
r}\right)+\psi_2\left(t\right)\phi_2\left({\bf r}\right),
\label{eq:gpe2m}
\end{equation}
where the Wannier-like functions
\begin{equation}
\phi_{1,2}\left({\bf
r}\right)=\frac{1}{\sqrt{2}}\left[\phi_{gs}\left({\bf r}\right)\pm
\phi_{exc}\left({\bf r}\right)\right] \label{eq:psi12}
\end{equation}
are mostly localized in each well and orthogonal to each other.

By substituting Eq.~(\ref{eq:gpe2m}) in Eq.~(\ref{eq:gpe}) and then projecting onto the lowest states, two coupled dynamical equations for the amplitudes $\psi_{1,2}(t)$ are obtained of the form:
\begin{multline}
i\hbar\dot{\psi}_j=\left(\hbar\omega_j-i{\gamma}/{2}\right)\psi_j+g\,|\psi_j|^2\psi_j
-J\psi_{3-j}
\\
+b\left[\left(2|\psi_j|^2+|\psi_{3-j}|^2\right)\psi_{3-j}+\psi_{3-j}^*\psi_j^2\right]\\
+c\left(2|\psi_{3-j}|^2\psi_j+\psi^*_{j}\psi^2_{3-j}\right)+F_j(t).\label{eq:gpe2mod}
\end{multline}
Linear dynamics is summarized by the diagonal term
\begin{eqnarray}
\hbar\omega_j=
\frac{1}{2}\left(E_{gs}+E_{exc}\right)\,, \label{eq:omega}
\end{eqnarray}
and the linear hopping coefficient
\begin{equation}
J=\frac{1}{2}\left(E_{exc}-E_{gs}\right)\,.\label{eq:J}
\end{equation}
Pumping is described by
\begin{equation}
F_j(t)=\int d{\bf r}\,\phi_j({\bf r})\,f({\bf r},t)\,, \label{eq:Fpar}
\end{equation}
while nonlinear effects are described by the three coupling coefficients
\begin{eqnarray}
g&=&v\int d{\bf r} \left[\phi_{1,2}({\bf r})\right]^4 \label{eq:g} \\
b&=&v\int d{\bf r} \phi_{2,1}({\bf r})\left[\phi_{1,2}({\bf
r})\right]^3 \label{eq:b} \\
c&=&v\int d{\bf r} \left[\phi_{1}({\bf r})\right]^2
\left[\phi_{2}({\bf r})\right]^2\,, \label{eq:c}
\end{eqnarray}
For typical geometries and for moderate intensities, the
coefficients $b$ and $c$ are much smaller than the other
quantities and can be safely neglected. Within this
approximation, Eq.~(\ref{eq:gpe2mod}) reduces to the two mode
model Eqs. (\ref{eq:dyn2mod_1}-\ref{eq:dyn2mod_2}) used in the
previous Sections.

\subsection{Comparison with the 2-mode model}

In order to verify the validity of the two-mode approximation, a
numerical integration of the full GPE Eq.~(\ref{eq:gpe}) can be
performed and then quantitatively compared to the predictions of
the two-mode model.

For this comparison, realistic parameters for typical polariton
boxes in GaAs based microcavities \cite{eldaif06,kaitouni06} are
used, that is a trapping potential depth $U_0=5~\mbox{meV}$,
lateral box sizes $L_x=1~\mu\mbox{m}$ and $L_y=2~\mu\mbox{m}$, a
polariton mass $m_p=7\times10^{-5}m_0$ ($m_0$ being the electron
mass) and a nonlinear coupling constant $v=2 \times
10^{-3}~\mbox{meV}\mu\mbox{m}^{2}$. We assume the separation
$\delta=0.5~\mu\mbox{m}$ between the two wells. The corresponding
potential $U_{ext}$ and the functions relative to the first two
GP modes are displayed in Fig.~\ref{fig:potpsi}. A spatial width
$\sigma=0.5~\mu\mbox{m}$ is taken for the pump spot.

\begin{figure}[ht]
\includegraphics[width=.43 \textwidth]{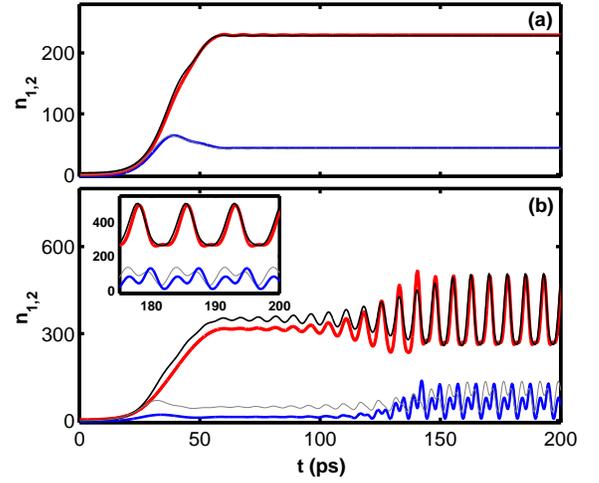}
\caption{Dynamical evolution of $n_1$ and $n_2$ ($n_2>n_1$), as
obtained from the full numerical calculation of GPE (thin lines)
and from the two-mode model (thick lines). Panel (a): GPE
simulations with pump energy $\hbar\omega=-2.8~\mbox{meV}$ and
pump amplitude $f_0=5~\mbox{meV}$; parameters of the
corresponding effective two-mode model:
$\hbar\omega_{1,2}=-3.15~\mbox{meV}$, $J=0.45~\mbox{meV}$,
$g=10^{-3}~\mbox{meV}$ and $F_1=5.8~\mbox{meV}$. Panel (b): GPE
simulations with pump energy $\hbar\omega=-2.8~\mbox{meV}$ and
pump amplitude $f_0=7.5~\mbox{meV}$; parameters of the
corresponding effective two-mode model:
$\hbar\omega_j=-3.1~\mbox{meV}$, $J=0.5~\mbox{meV}$,
$g=10^{-3}~\mbox{meV}$ and $F_1=8.5~\mbox{meV}$. }
\label{fig:confGPE}
\end{figure}

The time-evolution of the polariton occupation
$n_{1,2}(t)=|\psi_{1,2}(t)|^2$ in the two wells is plotted in
Fig.~\ref{fig:confGPE}. In the present GPE framework, the
occupations $n_{1,2}$ of each of the two wells are defined by the
spatial integrals
\begin{eqnarray}
n_1(t)=\int_{0}^{+\infty}\!\!\!\!\!dx\int\! dy\,\,|\Psi({\bf r},t)|^2  \\
n_2(t)=\int_{-\infty}^0\!\!\!\!\!dx \int \! dy\,\,|\Psi({\bf r},t)|^2.
\end{eqnarray}
Two different parameter choices are made in the two panels of
Fig.~\ref{fig:confGPE}. In (a), the system tends to a stable
stationary state, while self-pulsing oscillations are visible in
(b). The qualitative agreement with respectively panels (a,b) and
(c) of Fig.~\ref{fig:TIMEFcont} is apparent. Note that the
switch-on time of the pump considered in Fig.~\ref{fig:confGPE} is
not long enough to guarantee a quasi-static evolution of the
system even in the unstable region: differently from
Fig.~\ref{fig:TIMEFcont}(c), the self-pulsing oscillations are
immediately visible without the need of a perturbation seed.

As the linear coupling $J$ is very sensitive to the shape of the
wavefunctions in the barrier, interactions may affect it quite
significantly, spoiling the quantitative agreement with the
two-mode model. In order to get a good quantitative agreement in
all regimes, a better approximation can be adopted for the
localized wavefunctions using for $\phi_{gs}({\bf r})$ and
$\phi_{exc}({\bf r})$ the two lowest-energy solutions of the
time-independent, equilibrium Gross-Pitaevskii equation:
\begin{equation}
\left(-\frac{\hbar^2\nabla^2}{2m_p}+U_{ext}\left({\bf r}\right)+
v\,N\,\left| \phi\left({\bf r}\right)\right|^2\right)\,\phi
\left({\bf r}\right)= E^{(N)}\,{\phi}\left({\bf r}\right).
\label{eq:statgpe}
\end{equation}
The total polariton number $N=n_1+n_2$ has to be obtained from
the long-time limit $t\rightarrow \infty$ of the full GP equation
once the solution has come to their asymptotic steady state, or
by averaging over the period of the self-pulsing oscillations.
Correspondingly, Eqs. (\ref{eq:omega}) and (\ref{eq:J}) have to
be substituted by the N-dependent equations
\begin{eqnarray}
\hbar\omega_j&=&\frac{1}{2}\left(E_{gs}^{(N)}+E_{exc}^{(N)}\right)+\nonumber\\
&&-\frac{vN}{2}\int d{\bf r}\left([\phi_{gs}({\bf
r})]^4+[\phi_{exc}({\bf r})]^4\right)\, \label{eq:omegaN}
\end{eqnarray}
and
\begin{equation}
J=\frac{1}{2}\left(E_{exc}^{(N)}-E_{gs}^{(N)}\right)\,,\label{eq:JN}
\end{equation}
respectively. As expected, the main effect of interactions is to
slightly lift the bottom of the two wells, so to effectively
reduce the barrier height and enhance tunneling.

The result of such a procedure is also shown in
Fig.~\ref{fig:confGPE}: the overall qualitative agreement is good.
From a quantitative point of view, the agreement is always
excellent in the stable regime of panel (a), while some
discrepancies are visible in panel (b), in particular at short
times before the self-pulsing sets in. This can be expected, as
the parameters of the two-mode model extracted from the late time
dynamics slightly differ from what one would get from the early
stages. Furthermore, the spatial profile of the wavefunction has
a significant variation in time during the self-pulsing dynamics.

\section{Conclusions}
\label{par:conclu}

In this paper we have studied the two-mode dynamics of spatially coupled polariton boxes under a coherent external pumping and we have shown that this provides an interesting non-equilibrium optical generalization of the well-known Josephson effect of weakly coupled superfluids and superconductors.

For a continuous wave pumping, a phase diagram has been obtained which summarizes the steady state of the system as a function of pump intensity and frequency.
Stable and unstable regions have been identified; one-mode and parametric instabilities have been shown to be intrinsecally related to respectively optical bistability and self-pulsing effects. The response of the system to an additional probe provides unambiguous information on the Bogoliubov modes around the stationary state.

For a short pump pulse, the crossover from a Josephson oscillations regime to a self-trapping one has been characterized as a function of the pump intensity.
Peculiar features due to the non-equilibrium nature of the system have been pointed out.

The validity of the two-mode model and the actual observability of the predicted effects has been verified on the basis of the non-equilibrium Gross-Pitaevskii equation for polaritons in a double-well trap potential using parameters inspired by recent experiments.

D. S. and V. S. acknowledge financial support from the Swiss
National Foundation through Project No. PP002-110640.
IC is grateful to C. Ciuti, A. Recati, and A. Trombettoni for continuous stimulating discussions.

\end{document}